\documentclass[10pt,a4paper,twoside]{article}
\usepackage{graphicx}
\usepackage{psfrag}

\renewcommand{\vec}[1]{\mathbf #1}
\newcommand{\vecg}[1]{\mbox{\boldmath$#1$\unboldmath}}
\newcommand{\cc}[1]{#1^*}
\newcommand{\eff}{\mathrm{eff}}
\newcommand{\expect}[1]{\langle #1 \rangle}

\renewcommand{\d}{\mathrm{d}}
\newcommand{\rmi}{\mathrm{i}}
\renewcommand{\Re}{\mathrm{\,Re\,}}

\newcommand{\text}[1]{\mathrm{#1}}
\newcommand{\Hetrimer}{$^4$He$_3$}
\newcommand{\Hedimer}{$^4$He$_2$}

\newcommand{\phig}{\phi_{\gamma}}
\newcommand{\phigp}{\phi_{\gamma'}}
\newcommand{\Eg}{E_{\gamma}}
\newcommand{\Egp}{E_{\gamma'}}

\newcommand{\DD}{D$_2$}
\newcommand{\oDD}{o-D$_2$}

\newcommand{\HHDD}{H$_2$D$_2$}

\newcommand{\DDDD}{(\DD)$_2$}
\newcommand{\oDDDD}{(o-\DD)$_2$}

\newcommand{\pp}{\text{PP}}
\newcommand{\ttredpp}[1]{t_{#1}^\pp}
\newcommand{\tred}{t}
\newcommand{\taupp}[1]{\tau_{#1}^\pp}
\newcommand{\taudimggp}{\tau_{\gamma\gamma'}^\text{dim}}
\newcommand{\tautriggp}{\tau_{\gamma\gamma'}^\text{tri}}
\newcommand{\trans}[2]{#1$\rightarrow$#2}

\title{Diffraction of Weakly Bound Clusters: Spectroscopy and Size
  Effects\footnote{Invited talk given at the TH2002 International
    Conference on Theoretical Physics, Paris, July 22-27 2002}}

\author{
Martin~Stoll\footnote{Institut f\"ur Theoretische Physik, Universit\"at G\"ottingen}, 
Thorsten~K\"ohler\footnote{Clarendon Laboratory, Department of Physics, University of
  Oxford}, 
and Gerhard~C.~Hegerfeldt\addtocounter{footnote}{-1}\footnotemark[\value{footnote}]\addtocounter{footnote}{1}
}

\begin{document}

\maketitle

\begin{abstract}
  Exciting experiments in the field of atom and molecule optics have
  lately drawn much attention to the effects involved in the coherent
  diffraction of particle beams. We review the influence of the finite
  size of the particles and of their energy level spectrum on the
  diffraction pattern. In turn, we demonstrate how experimental
  diffraction measurements allow to determine these quantities of
  weakly bound molecules by considering the diffraction of \Hedimer,
  \Hetrimer, and \DDDD.
\end{abstract}

\section{Introduction}
As early as 1927 the wave-like properties of matter were demonstrated
when an electron beam was diffracted by a crystal of nickel
\cite{DG_PR30}. Recently, propelled by the availability of custom-made
nano-structured transmission gratings with periods as small as 100 nm,
typical diffraction experiments from light optics have been carried
over to the domain of atom and molecule optics
\cite{LB_AJP37,KSSP_PRL61,CM_PRL66,ST_SCIENCE266}. Nowadays, these
experiments serve as precision measurement devices for, e.g., the
interaction of atoms and molecules with solid surfaces
\cite{GSTHK_PRL83,BFGTHKSW_EPL59} or as quantum mechanical mass
spectrometers \cite{CEHRSWP_PRL74,GSTHKS_PRL85}.

Owing to the great sensitivity of these experiments it was soon
realized that a semi-classical theory using de Broglie waves and the
results from light optics otherwise does not yield an accurate
explanation of the experimental observations
\cite{GSTHK_PRL83,GSTHKS_PRL85}. Only by a fully quantum mechanical
approach, based on few-body scattering theory, could the diffraction
pattern of a collimated beam of helium atoms and helium clusters be
described \cite{HK_PRA57, HK_PRA61}. Since all particles in such a
beam share the same (average) velocity their de Broglie wave lengths
are inversely proportional to their masses. Therefore, their
diffraction angles are scaled by the same factor. On the one hand this
mass-selective property allowed to ascribe certain experimental
diffraction maxima to heavy clusters up to $^4$He$_{26}$
\cite{BST_JCP117}.  On the other hand the diffraction of the
diatomic helium cluster (``dimer'') could be quantitatively studied
and its bond length and binding energy could be determined
\cite{GSTHKS_PRL85}.

Moreover, atoms or molecules may coherently undergo internal
transitions upon interaction with the diffracting object. Inelastic
diffraction from a transmission grating involving transitions of
metastable argon atoms \cite{BBD_EPJD17} and nitrogen dimers
\cite{BBP_EPL56} has already been observed. In these experiments,
however, the grating served only to increase the intensity diffracted
by the slits. In the case of coherent diffraction from many slits of a
grating, inelastic diffraction gives rise to separate additional
diffraction maxima. The diffraction angles at which these new maxima
appear are related to the transition energies in a characteristic way.
This was first discussed for the particularly interesting case of the
three-body cluster of helium (``trimer'') \cite{HK_PRL84} in order to
produce the predicted excited state, which is believed to be of
Efimov-type (see, e.~g., Ref.~\cite{ELG_PRA54}).  Recently, the theory
of inelastic diffraction was also applied to the examples of the van
der Waals dimers \HHDD\ and \DDDD\ \cite{SK_JPB35}. Their energy level
spectra are well known \cite{D_JPB16}. Therefore, they may serve to
compare the theory to experimental data.

\section{Diffraction of very weakly bound molecules}

In this section we will review some aspects of the theory of molecular
diffraction. Our approach
has been described in detail before \cite{HK_PRA57, HK_PRA61}.
Therefore, we will focus on the final steps of the calculation which
lead to the desired results. These results relate the experimentally
recordable diffraction pattern to internal quantities of the molecular
bound state: the energy level spectrum and the bond length. Only the
experimentally most relevant case of diffraction from a nano-scale
transmission grating will be considered.

\begin{figure}[htbp]
  \begin{center}
    \includegraphics[width=0.6\textwidth]{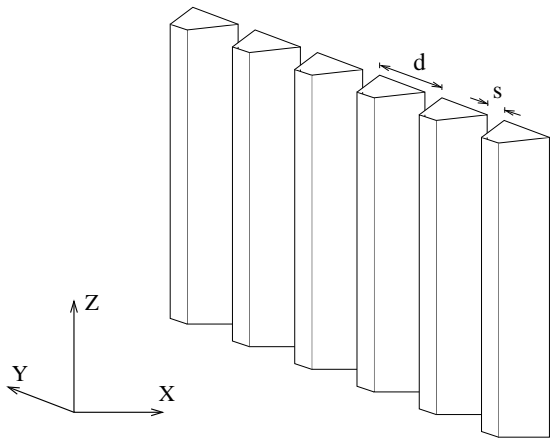}
    \caption{The coordinate system is chosen such that the bars of the
      transmission grating are parallel to the $z$ axis, and that the
      grating is periodic in the $y$ direction. The period is denoted
      by $d$, and the slit width by $s$.}
    \label{fig:grating}
  \end{center}
\end{figure}

We choose the coordinate system such that the $z$ axis is parallel to
the bars of the grating (cf.~figure \ref{fig:grating}). 
As an idealization, we assume the grating to be translationally
invariant along the $z$ axis. Therefore, the $z$-component of the
incident momentum is conserved, and the diffraction problem has
effectively been reduced to the $xy$ plane.  
A typical transmission grating, as considered in this article, has a period of
$d=100$ nm and a slit width of $s\approx 70$ nm. The de Broglie wave length of
a helium dimer at a velocity of 1~km/s, for example, is
$\lambda_\text{dB}\approx 0.05$~nm. Therefore, the diffraction angles are of
the order of several mrad about the forward direction.

The diffraction pattern of a well-collimated particle beam is, quite
generally, determined by the transition amplitude between the incident
and final momenta. For atoms, which are regarded as point-like
particles (PP) here, of mass $M$ with incident momentum $\vec P'$ and final
momentum $\vec P=\vec P'+\Delta\vec P$, the transition amplitude is
given by \cite{HK_PRA61}
\begin{eqnarray}
  \nonumber
  \lefteqn{\ttredpp{}(P_{x},P_{y}; P'_{x},P'_{y})=}& &\\
  \label{eq:ttredpp}
  & &
  -\rmi\frac{P_{x}}{(2\pi)^2 M\hbar}
  \int\d Y\,
  \exp(-\rmi\Delta P_{y} Y/\hbar)
  \left[1-\taupp{}(P'_{x},P'_{y};Y)\right].
\end{eqnarray}
Here, $\taupp{}$ is the transmission function characterizing the
grating. In the special case where the grating bars are assumed to be
purely repulsive, equation (\ref{eq:ttredpp}) recovers the well-known
Kirchhoff diffraction amplitude from classical optics \cite[chap.
8.5]{bornwolf}.

\begin{figure}[htbp]
  \begin{center}
    \includegraphics[width=0.8\textwidth]{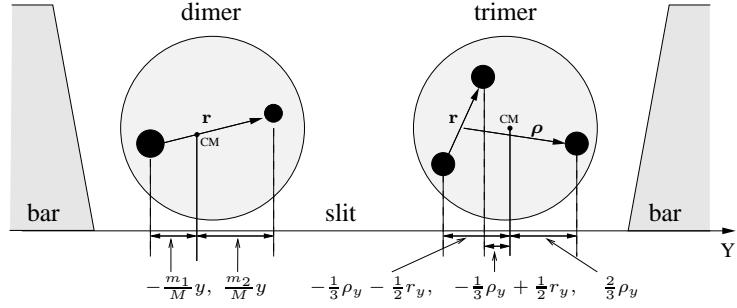}
    \caption{The position arguments of the functions $\taupp{i}$ in
      equations (\ref{eq:taudimggp}) and (\ref{eq:tautriggp}) can be
      interpreted as projected displacements of the positions of the
      constituents of the dimer, or trimer, from the center of mass position
      $Y$.}
    \label{fig:tridiffr}
  \end{center}
\end{figure}

A corresponding result may be obtained for dimers. The masses of the
constituents of a dimer are denoted by $m_1$ and $m_2$, and the total mass is
$M=m_1+m_2$. The interaction between the constituents is accounted for by a
two-body potential $V(\vec r)$, where $\vec r$ denotes the relative coordinate
in the dimer. Bound dimer states supported by $V(\vec r)$ are denoted by their
wave functions $\phig(\vec r)$ with corresponding binding energies $\Eg$. In
the following, we will always assume that the kinetic energy of an incident
dimer exceeds the ground state energy $E_0$ of the dimer by far.
This requirement is easily met for the cases considered in this
article. For example, the kinetic energy of a helium dimer at a
velocity of 1 km/s is of the order of 40~meV whereas the binding
energy of its single bound state is $|E_0|\approx0.1 \mu$eV
\cite{GSTHKS_PRL85}. Hence upon diffraction from the grating,
excitation as well as break-up of the dimer may occur. For an incident
dimer in bound state $\phigp$ with momentum $\vec P'$, the transition
amplitude to the state $\phig$ with final momentum $\vec P=\vec
P'+\Delta\vec P$ is given by \cite{HK_PRA61}
\begin{eqnarray}
  \nonumber
  \lefteqn{\tred(P_x,P_y,\phig;P'_x,P'_y,\phigp)=}& &\\
  \label{eq:treddim2}
  & &
  -\rmi\frac{P_{x}}{(2\pi)^2 M\hbar}
  \int\d Y\,
  \exp(-\rmi\Delta P_y Y/\hbar)
  \left[\delta_{\gamma\gamma'}-\taudimggp(P'_x,P'_y;Y)\right],
\end{eqnarray}
where the corresponding dimer transmission function is given as the
weighted product of two point-particle transmission functions (cf.~figure
\ref{fig:tridiffr}),
\begin{eqnarray}
  \label{eq:taudimggp}
  \lefteqn{\taudimggp(P'_x,P'_y;Y)=\int\d^3r\ 
  \cc{\phig}(\vec r)\phigp(\vec r)}& &\\
  \nonumber
  & &\times \
  \taupp{1}\left(\frac{m_1}{M}P'_x,\frac{m_1}{M}P'_y; 
    Y+\frac{m_2}{M}y\right)
  \taupp{2}\left(\frac{m_2}{M}P'_x,\frac{m_2}{M}P'_y; 
    Y-\frac{m_1}{M}y\right).
\end{eqnarray}

Finally, under similar assumptions the transition amplitude for
trimers may be derived. For simplicity, we specialize to the case
where all three constituents have equal masses $m=M/3$, as in the
helium trimer \Hetrimer. It can be shown that equation
(\ref{eq:treddim2}) also holds for trimers, if the dimer transmission
function is substituted by its trimer equivalent,
\begin{eqnarray}
  \nonumber
  \tautriggp(P'_x,P'_y;Y)=\int\d^3r\ \d^3\rho\ 
  \cc{\phig}(\vec r,\vecg\rho)\phigp(\vec r,\vecg\rho)\ 
  \taupp{1}\left(\frac13 P'_x,\frac13 P'_y; 
    Y+\frac23 \rho_y\right) \\
  \label{eq:tautriggp}
  \times
  \taupp{2}\left(\frac13 P'_x,\frac13 P'_y; 
    Y-\frac13\rho_y+\frac12 r_y \right)
  \taupp{3}\left(\frac13 P'_x,\frac13 P'_y; 
    Y-\frac13\rho_y-\frac12 r_y\right).
\end{eqnarray}
The Jacobi coordinates $(\vec r, \vecg\rho)$, and an interpretation of the
position arguments involving $r_y$ and $\rho_y$ are shown in figure
\ref{fig:tridiffr}.
\psfrag{psft5}{\scriptsize $\frac{m_2}{M}y$}
\psfrag{psft6}{\scriptsize $-\frac{m_1}{M}y,$}
\psfrag{psft1}{\scriptsize $-\frac13\rho_y-\frac12r_y,$}
\psfrag{psft2}{\scriptsize $-\frac13\rho_y+\frac12r_y,$}
\psfrag{psft3}{\scriptsize $\frac23\rho_y$}
\psfrag{psftr}{\scriptsize $\vec r$}
\psfrag{psftrho}{\scriptsize $\vecg \rho$}

Whereas the intensities of the diffraction maxima are given by the
modulus square of the transition amplitude \cite{HK_PRA61}, two
kinematic relations determine their angular positions.  Firstly,
energy conservation requires that
\begin{equation}
\label{eq:erergcons}
(P_x'^2+P_y'^2)/2M+\Egp=(P_x^2+P_y^2)/2M+\Eg
\end{equation}
(for the case of atom diffraction the binding energies $\Egp, \Eg$
should be set to zero). Secondly, the discrete periodicity of the
grating implies the conservation of the lateral momentum $P'_y$ up to
a reciprocal lattice vector, i.e.
\begin{equation}
  \label{eq:DeltaP2}
  \Delta P_y=n2\pi\hbar/d\ ,\quad n=0,\pm 1, \pm2,\ldots.
\end{equation}
For atoms this requirement can easily be derived from equation
(\ref{eq:ttredpp}) if the transmission function
$\taupp{}(P'_{x},P'_{y};Y)$ is assumed to have period $d$ in its third
argument. An analogous derivation can be carried out for dimers and
trimers using equation (\ref{eq:treddim2}).

Introducing polar coordinates in the $xy$ plane, a short calculation
with equations (\ref{eq:erergcons}) and (\ref{eq:DeltaP2}) shows that
the $n$-th order principal diffraction maximum is located at
\cite{HK_PRL84,SK_JPB35}
\begin{equation}
  \label{eq:thetan}
  \sin\theta_n=\left(1-\frac{\Eg-\Egp}{P'^2/2M}\right)^{-1/2}
  \left[\sin\theta'+n\frac{2\pi\hbar}{P'd}\right].
\end{equation}
where $\theta_n$ is related to the final momentum by
$P_y=P\sin\theta_n$, and the incident angle $\theta'$ is defined by
$P'_y=P'\sin\theta'$. Two cases may be distinguished: for elastic
($\Eg=\Egp$) diffraction the square root factor in equation
(\ref{eq:thetan}) becomes unity and the classical wave optical result
is recovered. For inelastic ($\Eg\neq\Egp$) molecular diffraction the
zeroth order principal maximum is shifted to a larger (smaller) angle
in the case of excitation (de-excitation), and the angular spacing of
the diffraction maxima is increased (decreased). We note that the
derivation of equation (\ref{eq:thetan}) involved solely the two
general conservation laws (\ref{eq:erergcons}) and (\ref{eq:DeltaP2}).
It applies, therefore, not only to dimers but also to larger clusters.

\section{The energy spectrum of a weakly bound molecule}

Molecular diffraction allows, in principal, to determine the
transition energies between bound states of a weakly bound molecule.
The key is to note that, according to equation (\ref{eq:thetan}),
inelastic diffraction gives rise to additional diffraction maxima
whose experimentally measurable angular spacing depends upon the
transition energies $\Eg-\Egp$.

To demonstrate the results presented above we consider the deuterium
mole\-cule dimer in its ortho-ortho modification, \oDDDD. At low nozzle
temperatures both constituents may be assumed to be in their lowest
rotor states $(j=0)$ \cite{SK_JPB35}. A dimer of two such \oDD\ 
exhibits four rotational bound states, labeled by their end-over-end
rotational quantum number $l$. All four bound states belong to the
lowest vibrational mode \cite{D_JPB16}. The interaction with the
grating induces transitions between these dimer states. In the special
case of equal constituents, as in \oDDDD, however, the dimer
transmission function (\ref{eq:taudimggp}) vanishes due to symmetry if
the incoming $(l')$ and outgoing $(l)$ dimer states have opposite
parity.  This gives rise to the parity selection rule
$l'+l=\text{even}$.

\begin{figure}[htbp]
  \begin{center}
    \includegraphics[width=0.6\textwidth]{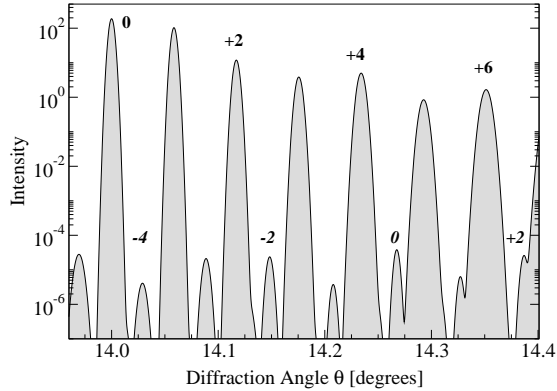}
    \caption{Calculated diffraction pattern of an \oDDDD\ beam at an
      angle of incidence of $\theta'=14^\circ$. In-between the intense
      elastic maxima the weaker and more widely spaced inelastic
      maxima due to the \trans{0}{2} transition are visible. Other
      inelastic transitions are not resolved.
      \label{fig:DDDD}}
  \end{center}
\end{figure}
The population density of the $l'$-states in the incident beam may be
estimated by their angular momentum degeneracy and the Boltzmann
weight factors $(2l'+1)\exp(-E_{l'}/k_B T_b)$, where the translational
beam temperature is assumed to be $T_b=400$mK \cite{SK_JPB35}.
Therefore, initially only the $l'=0$ ($E_0=-848\mu$eV) and $l'=1$
($E_1=-720\mu$eV) states are significantly populated \cite{SK_JPB35}.
Figure \ref{fig:DDDD} shows a diffraction pattern for \oDDDD\ at an
incident angle of $\theta'=14^\circ$. The intense maxima are due to
elastic diffraction whereas the weak maxima arise from the transition
\trans{0}{2}. The wider angular spacing of the latter is related to
the transition energy $E_2-E_0\approx377\mu$eV. Inelastic maxima of
other transitions are too weak to be resolved.

\section{The size of a weakly bound molecule}

In this section we will review how the relative intensities in an
elastic molecular diffraction pattern may be used to determine the
bond length of a molecule. Starting from the transition amplitudes
(\ref{eq:ttredpp}) or (\ref{eq:treddim2}) it has been shown before
\cite{GSTHK_PRL83,GSTHKS_PRL85} that the $n$-th order diffraction
maximum intensity is of the general form
\begin{equation}
  \label{eq:In}
  I_n \propto\ 
  \frac{\sin^2\left(\pi n s_\eff/d\right)+
    \sinh^2\left(\pi n \delta/d\right)}
  {(\pi n/d)^2}
  \ \exp\left[-(2\pi n\sigma/d)^2\right].
\end{equation}
This expression contains a Kirchhoff slit function whose effective
slit width $s_\eff$ is smaller than the geometrical slit width $s$ of
the grating. The difference accounts for the finite size of the
molecule and for its van der Waals interaction with the grating. (The
parameters $\delta$ and $\sigma$ in equation (\ref{eq:In}) are also
related to the van der Waals interaction, to irregularities of the
grating, and to dimer break-up, but they will not be relevant in the
following.) Generally, the effective slit width assumes the velocity
dependent form
\begin{equation}
  \label{eq:seffgen}
  s_\eff(v')=s-\Re \int_{-s/2}^{s/2}\d Y\ 
  \left[1-\tau(P'_x,P'_y;Y)\right],
  \quad v'=\sqrt{P'^2_x+P'^2_y}/M,
\end{equation}
where $\tau$ should be replaced by the respective transmission
function (cf.~equations \ref{eq:taudimggp} and \ref{eq:tautriggp}, and
Ref.~\cite{HK_PRA61} for atoms). Therefore, in the case of dimers, or
trimers, the full molecular bound state wave function enters into
equation (\ref{eq:seffgen}). The integration extends over one slit.

For the helium dimer, however, it was argued in
Ref.~\cite{GSTHKS_PRL85} that the dependence on the full wave function
may be neglected. Denoting the expectation value of the bond length by
$\expect{r}=\int\d^3r\ \left|\phi(\vec r)\right|^2 r$ one obtains,
after some algebra, the approximate expression
\begin{equation}
  \label{eq:seffdim}
  s_\eff(v')\approx s-\frac{\expect r}2 - 2 \Re \int_{0}^{s/2}\d Y\ 
  \left[1-
    \taupp{1}\left(
      Y\right)
    \taupp{2}\left(
      Y-\frac12 \expect{r}\right)
  \right],
\end{equation}
where the momentum arguments of the transmission functions have been
omitted for better readability.
\begin{figure}[htbp]
  \begin{center}
    \includegraphics[width=0.6\textwidth]{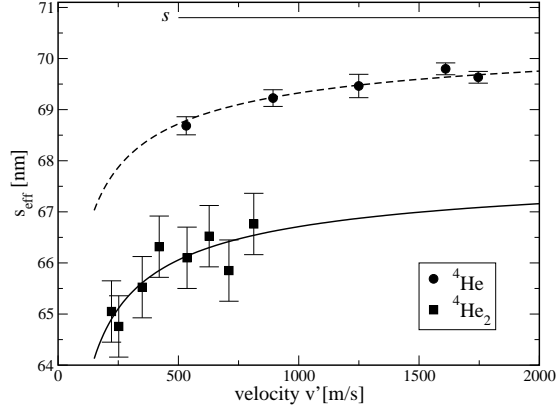}
    \caption{Experimental values for the effective slit widths of
      helium atoms (circles) and dimers (squares). The data were
      obtained by fitting equation (\ref{eq:In}) to experimental
      diffraction patterns measured by the Toennies group at the Max
      Planck Institut in G\"ottingen. The dashed line was calculated
      from equation (\ref{eq:seffgen}). The solid line was calculated
      from equation (\ref{eq:seffdim}) with $\expect{r}=5.2$ nm (best
      fit). The thin horizontal line at the top indicates the
      geometrical slit width $s$.}
    \label{fig:seffdim}
  \end{center}
\end{figure}
Figure \ref{fig:seffdim} shows helium atom (circles) and dimer (squares) data
for $s_\eff$, as determined by fitting equation (\ref{eq:In}) to experimental
diffraction patterns. The dashed line was calculated from equation
(\ref{eq:seffgen}) for helium atoms. The solid line was calculated from
equation (\ref{eq:seffdim}) with $\expect{r}$ adjusted such as to obtain the
best agreement with the data. A short analysis reveals that the integrals in
equations (\ref{eq:seffgen}) and (\ref{eq:seffdim}) vanish in the limit of
infinite velocity. Hence, the distance between the solid and the dashed lines
in figure \ref{fig:seffdim} allows to estimate the bond length to be
$\frac12\expect r\approx 2.5$ nm.
A quantitative evaluation of the data yields the final result
$\expect{r}=5.2\pm0.4$ nm, making the helium dimer the largest known
diatomic molecule in the ground state \cite{GSTHKS_PRL85}.

\begin{figure}[htbp]
  \begin{center}
    \includegraphics[width=0.6\textwidth]{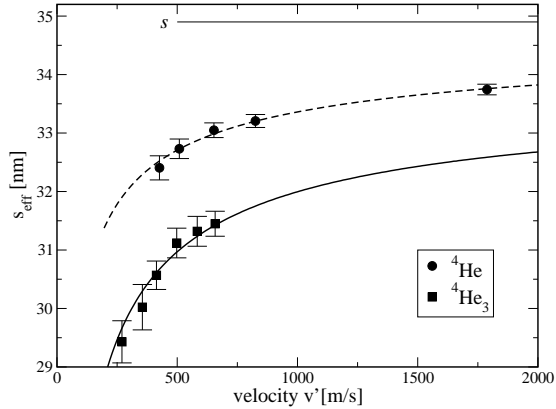}
    \caption{Experimental values for the effective slit widths of
      helium atoms (circles) and trimers (squares) at an incident
      angle of $\theta'=18^\circ$. The solid line was calculated from
      equation (\ref{eq:sefftri}) using the best fit value for
      $\expect r$. The thin horizontal line at the top indicates the
      projected slit width perpendicular to the incident beam.}
    \label{fig:sefftri}
  \end{center}
\end{figure}

For the helium trimer a similar analysis can be carried out. It
follows that the effective slit width can be expressed in terms of the
expectation value of the pair distance $\expect r=\int\d^3
r\int\d^3\rho\ \left|\phi(\vec r,\vecg\rho)\right|^2 r$ only. In this
way one obtains the approximate formula
\begin{eqnarray}
  \label{eq:sefftri}
  \lefteqn{s_\eff(v')\approx s-\frac34{\expect r}}& &\\
  \nonumber
  & & -2 \Re \int_{0}^{s/2}\d Y\ 
  \left[1-
    \taupp{1}\left(Y\right)\taupp{2}\left(Y-\frac12 \expect{r}\right)
    \taupp{3}\left(Y-\frac58 \expect{r}\right)
  \right].
\end{eqnarray}
To lessen the impact of the van der Waals interaction with the grating
in favor of a more accurate determination of the pair distance
$\expect r$ the measurements were carried out at an incident angle of
$\theta'=18^\circ$. Therefore, the geometrical slit width $s$ reduces
to its projection perpendicular to the incident beam. The data are
shown in figure \ref{fig:sefftri}.
Analogously to the helium dimer case a comparison between equations
(\ref{eq:sefftri}) and (\ref{eq:seffgen}) shows that the distance between the
dashed line (atoms) and the solid line (trimers) is asymptotically given by
$\frac34\expect r$. A detailed quantitative analysis of the pair distance is
yet underway.
\section{Conclusions}

A diffraction experiment setup is a sensitive measurement apparatus for weakly
bound molecules. We have summarized in this article how such a setup may be
employed to study two characteristic features of the bound state of a cluster:
The energy level spectrum and the bond length.

In the first part we demonstrated how the energy level spectrum of a
two-body cluster may be determined in a diffraction experiment by
measuring the angular positions of the inelastic maxima. The results
were applied to the realistic example of the diffraction of \oDDDD.
The derivation of the formula (\ref{eq:thetan}) for the inelastic
diffraction angles relies solely on two general conservation laws and
is, therefore, also applicable to larger clusters \cite{HK_PRL84}.

In the second part we showed in which way the experimental diffraction
intensities allow to determine the bond lengths of dimers and trimers.
The general and intuitive concept of an effective slit width was
presented, and applied to the clusters \Hedimer\ and \Hetrimer. A
comparison with experimental data from the group of J.P.~Toennies in
G\"ottingen was shown. Recently, in order to determine the bond length
of the helium trimer more accurately a non-zero incident angle was
chosen.  The analysis of this experiment is currently in progress.

\end{document}